\documentclass[aps,prb,reprint,amssymb,amsfonts]{revtex4-1} 
\usepackage{mathptmx} 
\usepackage{bm}
\usepackage{graphicx}
\usepackage{multirow}
\usepackage{color}
\usepackage{hyperref}
\hypersetup{
pdfborder={0 0 0}, 
colorlinks=true, 
linkcolor=blue,
citecolor=blue,
urlcolor=blue
}

\begin{document}

\title{Dissipation enhancement from a single vortex reconnection in superfluid helium}

\author{R. H\"anninen}
\affiliation{O.V. Lounasmaa Laboratory, Aalto University, P.O. BOX 15100, 00076 AALTO, Finland}

\date{September 16, 2013} 

\begin{abstract}
We investigate a single vortex reconnection event in superfluid helium at finite temperatures using the vortex filament model. The reconnection induces Kelvin waves which strongly increase energy dissipation. We evaluate the mutual friction dissipation from the reconnection and show that the dissipation power has universal form, which is seen by scaling both time (measured from the reconnection event) and power by the mutual friction parameter $\alpha$. This observation allows us to conclude that the Kelvin-wave cascade is not important in the energy dissipation process within the range $\alpha \gtrsim 10^{-3}$. Rather, the energy is directly transferred from Kelvin waves to the normal component. Moreover, while the excited Kelvin waves greatly enhance energy dissipation, no similar change is seen in angular momentum from the reconnection event. This result has similarities with recent $^3$He-B measurements. Our results also confirm another earlier observation that the minimum distance between vortices scales approximately as $d \propto \sqrt{|t-t_{\rm rec}|}$, both before and after the reconnection event.
\end{abstract}
%
\maketitle

\section{Introduction}\label{s.intro}

In quantum turbulence, energy dissipation in the zero temperature limit is a central open question.
\cite{Svistunov1995,VinenJLTP2002,BradleyPRL2006decay,FrontPRL2007,WalmsleyPRL2007} In the absence of
quantized vortices helium superfluids behave like an ideal Eulerian fluid: with zero viscosity, no turbulent
boundary layer is formed and the fluid exerts no drag on solid surfaces. The appearance of quantized
vortices breaks this ideal behavior and in many cases the superfluid then behaves quasi-classically, similar
to a classical viscous fluid. In classical fluids the energy dissipation arises due to the viscosity. In
quantum fluids vorticity is quantized in units of the circulation quantum. Here a hydrodynamic cascade, 
reminiscent of the Kolmogorov-Richardson cascade of classical hydrodynamics, can be expected to be effective 
only down to length scales of the order of the intervortex distance.\cite{BradleyPRL2006decay,WalmsleyPRL2007,NemirovskiiPRB2010} 
At smaller scales a Kelvin wave cascade on individual vortex lines is assumed to be the dominant dissipation 
mechanism, which transfers energy to scales where it can eventually be dissipated via microscopic processes.
\cite{Svistunov1995,KS2004,LN2010,SoninPRB2012} The existence and properties of the Kelvin wave cascade have 
been discussed for two decades, but no firm experimental proof has yet been provided.

Reconnections play an important role in the energy cascade, especially at length scales of the
order of the intervortex distance, where the energy is transferred from three-dimensional hydrodynamic
motion to one-dimensional Kelvin waves propagating along a line vortex.\cite{Svistunov1995}
In simulations a reconnection between two almost anti-parallel vortices has been shown to start a cascade of
vortex loop generation,\cite{Svistunov1995,KursaPRB2011,KerrPRL2011} which has been argued to be the dominant
dissipation mechanism for a sparse tangle in the low-temperature limit.\cite{KursaPRB2011,BradleyPRL2006decay}
In fact, evaporation of vortex loops from a localized vortex tangle has been suggested even more generally
to be responsible for the decay of turbulence at zero temperature.\cite{BarenghiPRL2002,NemirovskiiPRB2010,NemirovskiiPRB2012}
Analytically such a decay has been analyzed using a Gaussian model for vortex loops\cite{NemirovskiiPRB2010},
which resulted in a diffusion like equation for the vortex line density and which is supported by recent
simulations\cite{NemirovskiiPRB2012}.

At not too low temperatures mutual friction is the dominant dissipation mechanism, caused by the
scattering of quasiparticles from the vortex cores. At a reconnection event a vortex becomes strongly
distorted and locally large velocities appear on the curved vortex, giving rise to substantially increased
mutual friction dissipation. Previously the dissipation enhancement has only been evaluated for the viscous 
normal component using the coupled equations where the normal component is affected by the motion of the line 
vortex.\cite{KivotidesEPL2001}

These different possible mechanisms for energy dissipation in quantum turbulence motivated us to evaluate the
extra dissipation induced by a single reconnection event. We use the vortex filament model. It does not capture
the microscopics of the reconnection event on the length scale of the vortex core diameter. Instead our calculation 
examines the consequences from the reconnection event to the overall dynamics, by concentrating on larger length 
scales, i.e., those which extend from the typical inter-vortex distance two or three orders 
of magnitude lower to the resolution limit of our computation. We determine the increase in mutual friction 
dissipation after the reconnection event and find that a reconnection excites Kelvin waves whose mutual friction 
damping increases the direct energy loss. This loss appears to be roughly temperature independent. No formation 
of small new evaporating vortex loops is observed, nor any cascading of energy to smaller length scales. Finally, 
no large change, similar to the energy loss from the reconnection event, is seen in the dissipation of
angular momentum. These features appear to be consistent with recent experimental observations.\cite{HosioPRL2011,HosioNC2013}

The results cannot be directly generalized to the decay of dense vortex tangles. A single reconnection calculation
might overestimate dissipation in a tangle where reconnections are frequent and where there is not enough time
for mutual friction to dissipate all energy before the next reconnection occurs. This problem becomes more
pronounced at the lowest temperatures, where the decay time increases. Nevertheless, our calculations underline
the importance of the direct mutual friction losses in Kelvin wave excitations on approaching the zero-temperature
limit, which is of central current interest in the field of quantum turbulence.

\section{Model and equations}

In the vortex filament model\cite{schwarz85} all characteristic length scales
are assumed to be much larger than the vortex core radius, $a_0$, which in superfluid $^4$He is only of
order 0.1 nm. Within the vortex filament model the superfluid velocity, ${\bf v}_{\rm s}$, is given by the Biot-Savart law. At zero temperature the only force felt by the vortex is the classical Magnus force given by ${\bf f}_{\rm M} = \kappa\rho_{\rm s}({\bf v}_{\rm s}-{\bf v}_{\rm L})\times\hat{\bf s}'$ (per unit length), where ${\bf v}_L$ is the velocity of the vortex segment at point ${\bf s}$ on the vortex, with the unit vector $\hat{\bf s}'$ pointing in the tangential direction. Here $\kappa = \hbar/m_4$ = 0.0997 mm$^2$/s is the circulation quantum for superfluid $^4$He. Since the mass of the vortex core can typically be neglected, every point on the vortex simply moves at its local superfluid velocity:
\begin{equation}\label{e.bs2}
{\bf v}_{\rm s} =
\frac{\kappa}{4\pi}\hat{\bf s}'\times {\bf s}'' \ln\left(\frac{2\sqrt{l_{+}l_{-}}}{e^{1/2}a_0}\right) + 
\frac{\kappa}{4\pi}\int^{'}\frac{({\bf s}_1-{\bf s})\times d{\bf s}_1}
{\vert {\bf s}_1-{\bf s}\vert^3}\,. 
\end{equation}
Following Schwarz\cite{schwarz85}, the singularity in the Biot-Savart integral has here been removed, by extracting
out the local term (the first term on the right-hand side). 
Here $l_\pm$ are the lengths of the line segments connected to point ${\bf s}$  after discretization and ${\bf s}''$
denotes the normal at ${\bf s}$, where the derivation is with respect to the arc length $\xi$. In the second nonlocal
term the line integral is taken over the other vortex segments, not connected to ${\bf s}$.

At finite temperatures vortex motion is coupled via mutual friction to the motion of the normal component. The mutual
friction force (per unit length) acting on the vortex is given by
\begin{eqnarray}
{\bf f}_{\rm mf}/\rho_{\rm s}\kappa = \alpha \hat{\bf s}' \times \hat{\bf s}'\times ({\bf v}_{\rm s}-{\bf v}_{\rm n})
                 + \alpha' \hat{\bf s}' \times ({\bf v}_{\rm s}-{\bf v}_{\rm n}) \, ,
\label{e.fmf}
\end{eqnarray}
where $\alpha$ and $\alpha'$ are temperature-dependent and pressure-dependent mutual friction coefficients, whose
values are rather well known in superfluid $^4$He and $^3$He-B. (Note that the results for this paper
are calculated with $\alpha'$ = 0, which is approximately valid at low temperatures.) Thus at finite
temperatures the velocity of the vortex segment is given by
\begin{eqnarray}
{\bf v}_{\rm L} = {\bf v}_{\rm s} + \alpha \hat{\bf s}'\times({\bf v}_{\rm n}-{\bf v}_{\rm s})
-\alpha' \hat{\bf s}'\times \hat{\bf s}'\times({\bf v}_{\rm n}-{\bf v}_{\rm s}) \, .
\label{e.vL}
\end{eqnarray}
The power dissipated by mutual friction can be calculated from
\begin{eqnarray}
P_{\rm mf} = \int ({\bf v}_{\rm L} \cdot {\bf f}_{\rm mf}) d\xi = - \int ({\bf v}_{\rm L} \cdot {\bf f}_{\rm M}) d\xi \, .
\label{e.Pmf}
\end{eqnarray}
If we assume that the normal fluid is at rest, ${\bf v}_{\rm n} = 0$, then we obtain
\begin{eqnarray}
P_{\rm mf} = -\alpha\rho_{\rm s}\kappa \int |\hat{\bf s}' \times {\bf v}_{\rm s}|^2 d\xi \, .
\label{e.Pmf0}
\end{eqnarray}
The negative sign, which is typically omitted in the analysis below, indicates that energy is dissipated.
Since the mutual friction provides the only dissipation, which is considered here (in addition to numerical
dissipation), the power can also be calculated from the rate of change in the kinetic energy
of the superfluid component. However, at small $\alpha$ values fluctuations in the dissipation complicate the
calculation of the time derivative and we prefer to calculate $P_{\rm mf}$ from Eq.~(\ref{e.Pmf0}).
In calculating the energy ($E$), momentum ($P$), and angular momentum ($A$) we use the following line integrals\cite{SaffmanBook,Alekseenko2007}:
\begin{eqnarray}
E &=& \rho_{\rm s}\kappa \oint {\bf v}_{\rm s}\cdot {\bf s}\times \hat{\bf s}' d\xi \nonumber \\
{\bf P} &=& \frac{1}{2}\rho_{\rm s}\kappa \oint {\bf s}\times \hat{\bf s}' d\xi \label{e.EPA} \\
{\bf A} &=& \frac{1}{3}\rho_{\rm s}\kappa \oint {\bf s} \times {\bf s}\times \hat{\bf s}' d\xi \, . \nonumber
\end{eqnarray}

In the filament model a reconnection must be introduced by hand. However, its existence is supported by
Gross-Pitaevskii calculations\cite{KoplikLevinePRL1993} and recently also by experiments\cite{BewleyPNAS2008reconnection}.
Our method of making a reconnection follows that of other authors.
\cite{schwarz85,TsubotaPRB2000,KondaurovaJLTP2008,BaggaleyJLTP2012rec,KondaurovaLvov2013}
Traditionally one reconnects two vortices as soon as any two points approach closer than some given distance.
This critical distance is typically taken to be of the order of the resolution. We not only calculate the distances
between the different vortex points on a vortex, but also keep track of the minimum distance between two vortex segments.
(Between neighboring points the vortex is assumed to be straight, which is the standard assumption in the filament
model.) This distance can be smaller than the distance between two adjacent points on the same vortex. Thus this
reconnection procedure ensures good accuracy also in an adaptive point separation scheme, where the point separation is
adaptive, being much larger at locations where the local radius of curvature is large.

In these simulations the point separation is not adaptive, but is simply kept between
 $\Delta\xi_{\rm res}/2 < \Delta\xi < \Delta\xi_{\rm res}$, and a reconnection is made if any two segments get closer than
$0.4\,\Delta\xi_{\rm res}$, provided that the vortex length decreases in the reconnection event. Roughly speaking this
procedure ensures that the energy decreases. Numerically it might not always be evident that the energy $E$ decreases
in the reconnection (see, e.g., the peak in Fig.\,\ref{f.EPALatT0}), since the large curvatures and small distances complicate
the accurate calculation of $E$. The reason for avoiding an adaptive point discretization scheme, even if it would speed
up the calculation, is to minimize numerical errors: always when a point is added or removed the energy $E$ is changed
by a small amount. The point adjustment is done much more frequently in an adaptive scheme.
Clearly, the error could be eliminated by developing an energy conserving algorithm for adding or removing points.

\begin{figure}[!t]
\includegraphics[width=0.9\linewidth]{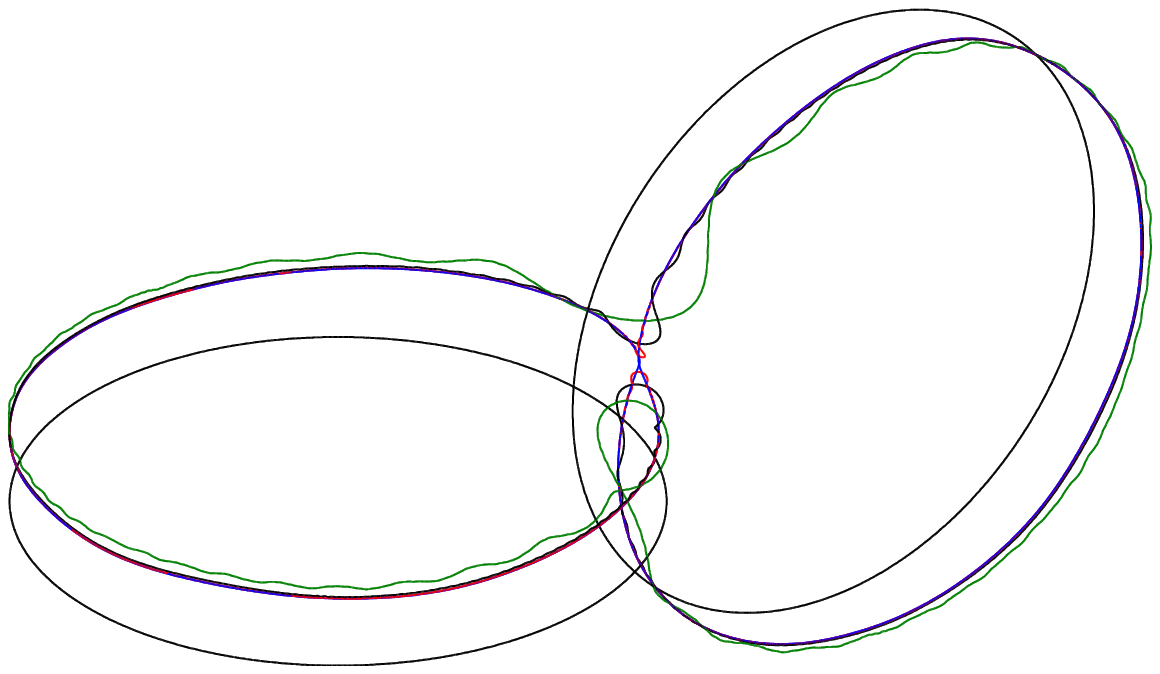} \\
\includegraphics[width=0.65\linewidth, angle=90]{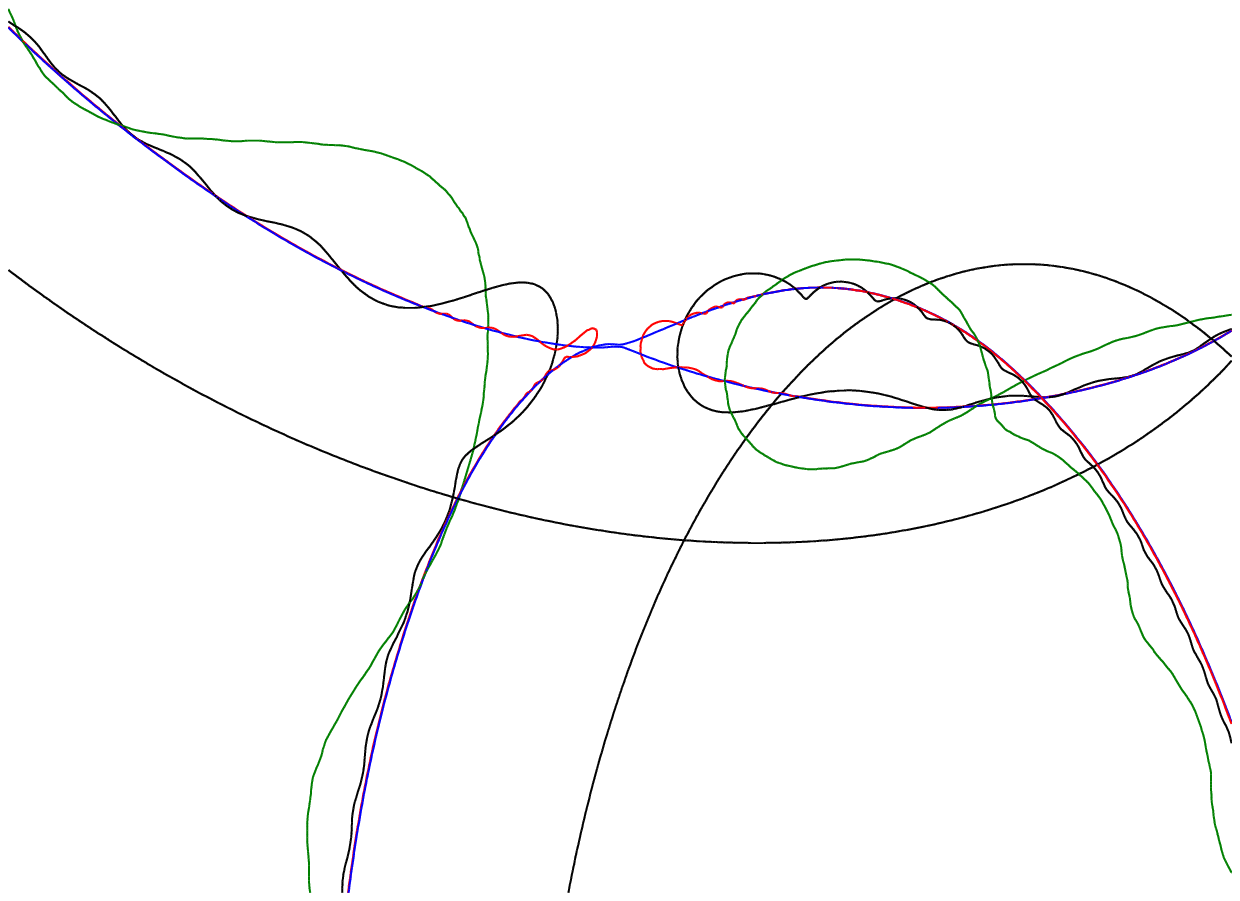}
\caption{(Color online)
Vortex reconnection at $T$ = 0. Configurations are for $t$ = 0 (black), 1.695
(blue, just before reconnection), 1.70 (red), 1.75 (black), and 2.00 (green) seconds.
\emph{Bottom panel:} The region near the reconnection site, zoomed and
rotated by 90 degrees.
}
\label{f.confsT0}
\end{figure}

One should note, however, that a numerical scheme, where the energy is too well conserved, might lead to a numerical
bottleneck near the resolution limit. This bottleneck would appear if the (Kelvin wave) cascade becomes important in
transferring energy from large scales to scales which cannot be resolved. In the calculation this would appear as a
fractalization of the vortex configuration, with the characteristic curvatures being of the order of the numerical
resolution. This is not the case here: even our zero-temperature calculations do not show any signs of accumulation of
the smallest scale structures. For instance, with a resolution $\Delta\xi_{\rm res} = 0.01\,$mm the calculation gives an
average curvature $\langle s'' \rangle \approx 8\,$mm$^{-1}$ which remains roughly constant from soon after the
reconnection up to 70\,s later. This we take as one of the indications that the Kelvin wave cascade is not triggered
by a single reconnection event, or that the cascade is too small to be observed (in accordance with Baggaley and
Barenghi\cite{BaggaleyPRB2011spectrum}, but in contrast to Kivotides \emph{et al.}\cite{KivotidesPRL2001}).

Finally, we note that even if we use $^4$He specific parameters, the results are also valid for superfluid $^3$He-B.
The slightly different value for the circulation and different core size can only change the numerical values
by a factor that is close to unity.

\section{Results}\label{s.results}

\begin{figure}[!t]
\includegraphics[width=0.98\linewidth]{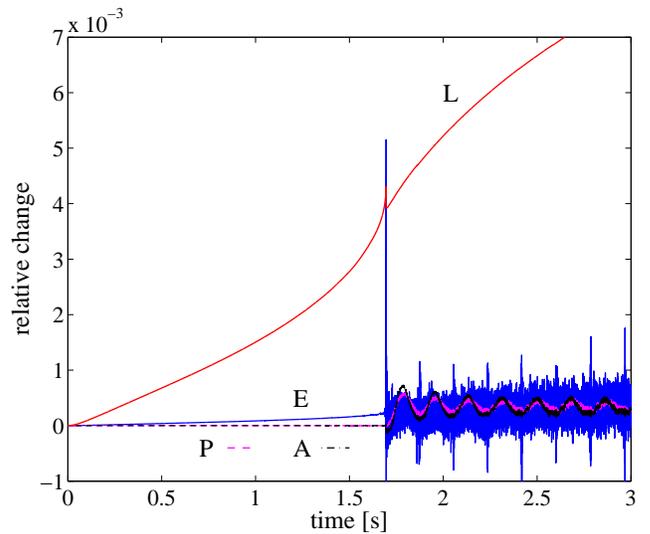}
\caption{(Color online)
Conservation of energy, $E$, momentum, $P$, and angular momentum, $A$, in our numerical calculation
at $T$ = 0, plus the evolution in vortex length, $L$.
The vortex reconnection occurs at $t$ = 1.6951 s, which is seen as a small drop in the vortex length,
together with a spike in the energy which is created by the numerical algorithm used for the reconnection.
The resolution is $\Delta\xi_{\rm res}$ = 0.0050\,mm.
}
\label{f.EPALatT0}
\end{figure}

Our starting configuration is two linked planar vortex rings, with radius $R$ = 1\,mm, separated by a
distance $d_0 = 0.1\,R$, and with their planes oriented perpendicular. The initial configuration, together
with some time development (at $T$ = 0, where $\alpha$ = $\alpha'$ = 0), is illustrated in Fig.~\ref{f.confsT0}.
The final configuration is one single closed loop. No additional small vortex rings are formed as a result of
the reconnection; i.e., no evaporation of extra vortex loops is observed. With closed loops, we can conveniently
evaluate the characteristics of the vortex configuration from the line integrals in Eqs.\,(\ref{e.EPA}).
Figure \ref{f.EPALatT0} illustrates that the energy and linear and angular momentum are all well conserved
in this calculation at $T$ = 0,
and that the typical numerical error is smaller than 0.1\,\%. This implies that our numerical error is well
smaller than the losses appearing in our finite-temperature simulations. Figure \ref{f.EPALatT0} also shows
that the vortex length is only an approximation for the energy and is not necessarily exactly constant at
zero temperature.

\begin{figure}[!t]
\centerline{
\includegraphics[width=0.98\linewidth]{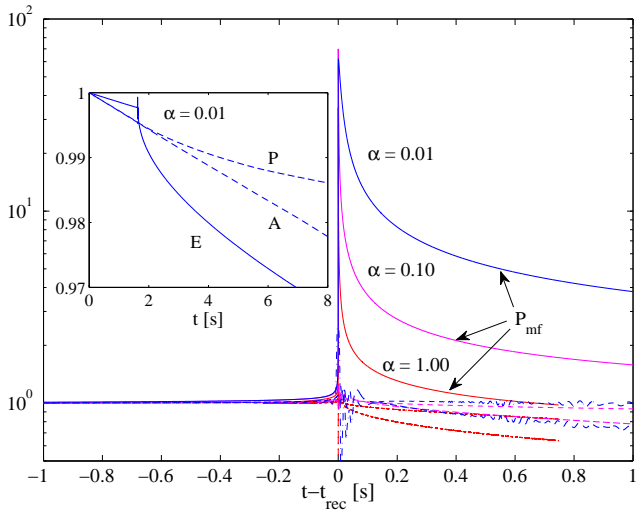}
}
\caption{(Color online)
\emph{Inset:} Time development of energy, $E$, momentum, $P$, and angular momentum, $A$, when $\alpha$ = 0.01
and $\alpha'$ = 0. The values plotted on the vertical axes have been divided by the initial values at $t$ = 0.
\emph{Main panel:} Energy dissipation rate, $P_{\rm mf} = dE/dt$ (solid lines), momentum $dP/dt$ (dashed), and
angular momentum $dA/dt$ (dash-dotted). The rates are scaled by their respective values at $t$ = 0. The three
different colors indicate the three different values of $\alpha$: 1 (red), 0.1 (magenta), and 0.01 (blue).
Note the logarithmic scale on the vertical axis.
}
\label{f.dissipEPA}
\end{figure}

The most important observation from the finite-temperature calculations with $\alpha > 0$ is
that energy dissipation is strongly increased by the Kelvin waves appearing immediately after
the reconnection, while momentum and angular momentum dissipation are barely affected. This is illustrated in
Fig.\,\ref{f.dissipEPA}, where the inset shows the time development of energy $E$,
momentum $P$, and angular momentum $A$, when $\alpha$ = 0.01. In the main panel we plot the
time derivatives $P_{\rm mf} = dE/dt$, $dP/dt$, and $dA/dt$ (scaled by their values at $t=0$), using various
values for $\alpha$. As seen here, only energy dissipation is strongly increased. The dissipation
of $A$ is barely affected by the reconnection and  the dissipation of $P$ is slightly reduced. The rapid
fluctuations near $t=t_{\rm rec}$ are likely numerical artifacts, since these features depend more sensitively
on the chosen resolution $\Delta\xi_{\rm res}$, while the overall behavior does not.

The explanation for the weak momentum dissipation is the following: The sharp kinks created in the
reconnection event induce Kelvin waves, but due to conservation of linear and angular momentum the Kelvin
waves on different sides of a kink must have opposite helicities (the $k$ vector of the Kelvin waves is
directed away from the kink). Even if the change in momentum is finite on both sides on the original kink,
the total change of momentum is zero because the changes have opposite signs. This is the case even in the
presence of finite mutual friction. However, both sides of the kink contribute equally
with positive signs to energy dissipation, which is therefore large compared to the total momentum dissipation.

\section{Analysis}\label{s.analysis}

The main problem in analyzing the reconnection in the vortex filament model is to extract the asymptotic behavior
before and after the event, which we do by subtracting a ``background'' dissipation power to which the reconnection
does not contribute. Well before reconnection, the dissipation is caused by the shrinking of the two independent
vortex rings. Here analytical estimation gives a good approximation. However, immediately after the reconnection
the situation is more complex. The late time configuration is close to a single vortex ring, with a larger radius
compared to the value of the initial rings, but which eventually shrinks away. Note that during these simulations
the total length has dropped 20\,\% or less. In the following, to extract the extra dissipation from the reconnection,
we make use of the fact that the time scale related to the simple shrinking of the two initial rings (or the final
perturbed ring) is much longer than the time scales related to the decay of the small scale structures (Kelvin waves)
which are induced by the reconnection event.

\begin{figure}[!t]
\centerline{
\includegraphics[width=0.98\linewidth]{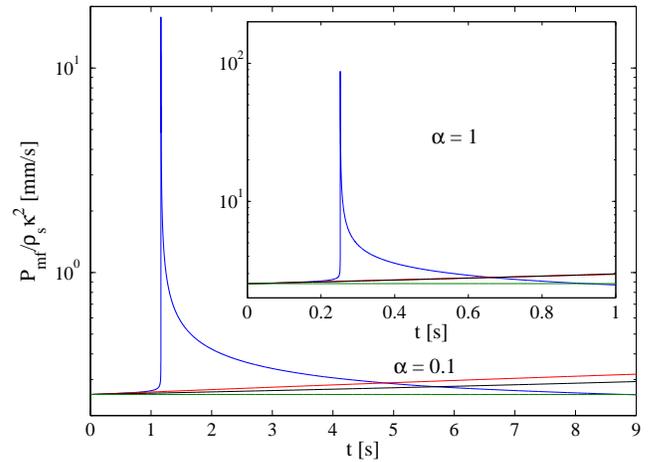}
}
\caption{(Color online)
Total mutual friction dissipation (blue) together with three different estimates for
the laminar background subtraction: linear extrapolation (red, largest), analytical calculation of
two independent rings (black), and constant approximation (green, smallest).
The \emph{main panel} is calculated for $\alpha$ = 0.1, and the \emph{inset} for $\alpha$ = 1.
}
\label{f.Pmfsample}
\end{figure}

In evaluating the extra dissipation we compare three different methods to estimate the ``background'' subtraction.
All of them overestimate the ``true'' background dissipation. The largest estimate is derived from a linear
extrapolation before the reconnection. The second estimate comes from an analytical calculation of two independent
vortex rings with a small adjustment of order 3\,\%  to correctly capture the initial dissipation. The third and
smallest estimate for the background subtraction is simply constant dissipation taken from the initial value,
$P_{\rm mf}(0)$. These estimates are plotted in Fig.~\ref{f.Pmfsample}. They provide a lower limit for the additional
dissipation from the reconnection because at later times the dissipation drops in all three cases below these
estimates. Also, the finite resolution underestimates the dissipation in the vicinity of the reconnection event,
where the maximum curvatures are of the order of our resolution.

Our simulations with different $\alpha$ values indicate that the power $P_{\rm mf}(t)$ takes a rather universal
form, which is obtained by scaling time with $\alpha$ and $P_{\rm mf}$ with $1/\alpha$. In other words, the
mutual friction power seems to be given by $P_{\rm mf} = \alpha f(\alpha{t})$, with some function $f(x)$.
The prefactor $\alpha$ comes quite naturally from Eq.~(\ref{e.Pmf0}), but the remaining dependence on
only $\alpha{t}$ must be approximate and can be attributed to the decay of independent Kelvin waves whose
amplitudes decay exponentially with a time scale proportional to $1/\alpha$.

\begin{figure}[!tb]
\centerline{
\includegraphics[width=0.98\linewidth]{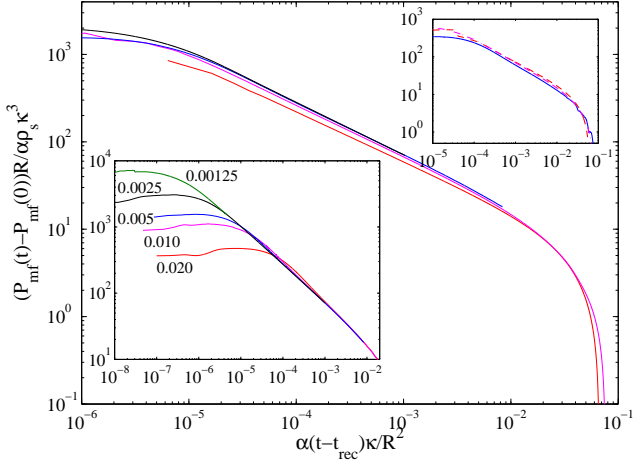}
}
\caption{(Color online)
Scaled additional mutual friction power from the reconnection event, plotted as a function of scaled time
$\alpha{t}$ after the reconnection. \emph{Main panel:} The results when $R$ = 1 mm at four
different values of $\alpha$: 1 (red), 0.1 (magenta), 0.01 (blue), and 0.001 (black). The resolution of the
calculations is $\Delta\xi_{\rm res} = 0.005\,$mm. \emph{Lower inset:} The extra dissipation at $\alpha$ = 0.01
is plotted with five different resolutions $\Delta\xi_{\rm res}$, which are denoted on the individual curves.
\emph{Upper inset:} The results with three different initial ring radii: $R$ = 0.1 mm (blue, solid line),
$R$ = 1.0 mm (magenta, dash-dotted), and $R$ = 10 mm (red, dashed), for which, respectively, the mutual friction damping
is $\alpha$ = 0.01, 0.10, and 1.00. The resolution is $\Delta\xi_{\rm res}/R$ = 0.020.
}
\label{f.Pmf0}
\end{figure}

This approximate scaling is illustrated in Fig.~\ref{f.Pmf0}, where we plot the scaled additional dissipation
using several values of $\alpha$ and different resolutions. In the lower inset the
extra dissipation is plotted for $\alpha = 0.01$ at five different resolutions.
Initially following the reconnection we see a maximum developing which increases steadily with resolution. However,
the integrated value of this dissipation contribution is almost invariant of the resolution, because later at intermediate
times the higher resolution curves drop below the lower resolution results, as seen in the inset. The asymptotics at late
times are very similar for all resolutions. Well after the reconnection there exists a wide region where the extra
dissipation power takes the form $f(x) = Ax^{-\gamma}$ with $\gamma \approx 0.6$. The late time behavior in the region,
where the curves bend strongly down, is sensitive to the approximation made for the background subtraction, but is
consistent with exponential damping of the longest Kelvin waves.

The scaling law for the mutual friction dissipation power, as verified in Fig.~\ref{f.Pmf0},
indicates that the Kelvin wave cascade
is not influential. Instead Kelvin waves at different wave lengths are excited directly by the reconnection, and
the energy is dissipated by these initially excited modes, and transferred to the normal component via mutual
friction damping. No cascade with nonlinear production of secondary modes is involved in a noticeable amount.
If the Kelvin wave cascade would start to dominate at some $\alpha$ value, then the cascade
would transfer energy to smaller scales, where it would be damped more effectively by mutual friction, breaking the
observed scaling law. Thus either direct mutual friction damping is still the most efficient dissipation mechanism
and cascade formation is too weak to be resolved, a single reconnection event is not sufficient to start the
cascade (as argued in Ref.~\onlinecite{BaggaleyPRB2011spectrum}), or finite-size effects limit the cascade,
such that it does not develop.

\begin{table}
\begin{tabular}{|c|c|c|c|c|c|c|c|}
$\alpha$ & $\Delta\xi_{\rm res}$ & $t_{\rm max}$ & $t_{\rm rec}$ & $E_{\rm rec}^{\rm linear}$ & $E_{\rm rec}^{\rm anal}$ & $E_{\rm rec}^{\rm const}$ & $E_{\tau=0.01}$\\
\hline
\multirow{3}{*}{1.0}                                               
  & 0.0200   & 1.0 & 0.24980  & 0.4857  & 0.4900  & 0.6072  & 0.1259 \\ 
  & 0.0100   & 1.0 & 0.25155  & 0.4820  & 0.4867  & 0.6041  & 0.1253 \\ 
  & 0.0050   & 1.0 & 0.25224  & 0.4819  & 0.4867  & 0.6042  & 0.1260 \\ 
\hline
\multirow{3}{*}{0.1}
  & 0.0200   & 10  & 1.15590  & 0.5172  & 0.5630  & 0.6573  & 0.1490 \\ 
  & 0.0100   & 10  & 1.16145  & 0.5088  & 0.5547  & 0.6489  & 0.1440 \\ 
  & 0.0050   & 9.0 & 1.16300  & 0.5161  & 0.5620  & 0.6562  & 0.1521 \\ 

\hline
\multirow{5}{*}{0.01}
  & 0.0200   & 100 & 1.61360  & 0.4345  & 0.6335  & 0.7123  & 0.1566 \\ 
  & 0.0100   & 85  & 1.62255  & 0.4363  & 0.6346  & 0.7131  & 0.1626 \\ 
  & 0.0050   & 10  & 1.62481  & -       &  -      &  -      & 0.1607 \\ 
  & 0.0025$^{1)}$ & 2.63  & 1.62512  & - &  -      &  -      & 0.1624 \\ 
  & 0.0013$^{2)}$ & 1.63  & 1.62519  & - &  -      &  -      & -      \\  
\hline
\multirow{3}{*}{0.001}
  & 0.0200   & 800 & 1.67860  & 0.2506  & 0.6294  & 0.7088  & 0.1491 \\ 
  & 0.0100   & 65  & 1.68638  & 0.2637  & -       & -       & 0.1627 \\ 
  & 0.0050   & 12  & 1.68813  & -       & -       & -       & 0.1641 \\ 
\hline
\multirow{2}{*}{0.000}
  & 0.0100   & 70  & 1.69323  & -  & -  & - & - \\
  & 0.0050   & 3.0  & 1.69514  & -  & -  & - & - \\
\hline
\end{tabular}
\caption{
Characteristic numbers from simulation runs at different $\alpha$ and different resolutions
$\Delta\xi_{\rm res}$. The initial configuration is two linked rings with $R$ = 1\,mm and a
separation of 0.1\,mm. Here $t_{\rm max}$ is the maximum time to which the calculation has been extended, while $t_{\rm rec}$
denotes the moment when the reconnection takes place. $E_{\rm rec}^{\rm appr}$ is the estimate
for the total extra dissipation from the reconnection, using the approximation
``\emph{appr}'' for the background subtraction (see text for details). The energy $E_\tau$,
defined in Eq.~(\ref{e.Ewindow}), is a measure of the extra energy dissipated during time
$\Delta{t}=\tau/\alpha$ after the reconnection, using the constant background subtraction.
All lengths are in millimeters and times are in seconds.
Energies are scaled by $\rho_{\rm s}\kappa^2$, which implies that they are in units of millimeters.
Notes: $^{1)}$ the starting configuration is taken from the lower resolution run
$\Delta\xi_{\rm res}$ = 0.0050 mm at $t$ = 1.60 s, while $^{2)}$ here it comes
from the run with $\Delta\xi_{\rm res}$ = 0.0025 mm at time $t$ = 1.62 s.
}
\label{t.recdata}
\end{table}

Table~\ref{t.recdata} provides a further summary of the results. It shows that the extra mutual friction dissipation
$E_{\rm rec}$ is rather independent of the resolution $\Delta\xi_{\rm res}$ and depends only weakly on $\alpha$. In fact,
the value of $E_{\rm rec}$ depends more sensitively on the background subtraction
than on $\alpha$.  The average for $E_{\rm rec}/\rho_{\rm s}\kappa^2$ is thus about 0.5 mm. The
energy per unit length for a straight vortex is $\rho_{\rm s}\kappa^2\ln(\ell/a)/(4\pi) \approx \rho_{\rm s}\kappa^2$,
where $\ell$ is some cut-off, typically the intervortex distance or the system size. Therefore, the dissipation
$E_{\rm rec}/\rho_{\rm s}\kappa^2$ corresponds to a reduction in length by about 0.5 mm. This is a sizable value.
For example, the estimation for the energy released by sound emission, using the Gross-Pitaevskii equation (GPE),
corresponds to a length reduction by a few coherence lengths only\cite{LeadbeaterPRL2001}. However, one
should note that these GPE simulations only apply at zero temperature.

At high resolution and small mutual friction it becomes numerically challenging to cover a sufficiently large
time window $\Delta{t}/\alpha$,  to capture all the relevant dissipation processes associated with the reconnection
event. To compare different simulations more quantitatively, we use a computationally more convenient time window,
which we define in terms of the following energy expression:
\begin{eqnarray}
E_\tau = \int_{t=0}^{\tau/\alpha}\lbrack P_{\rm mf}(t_{\rm rec}+t)-P_{\rm mf}(0) \rbrack dt \, .
\label{e.Ewindow}
\end{eqnarray}
This quantity is listed in the right most column of Table~\ref{t.recdata}. Here we have used the constant background
subtraction, although for $\tau \ll 1$ all three different approximations for the background subtraction give results
which differ only by a few percents. The total extra dissipated energy $E_{\rm rec}$ is obtained if the time integration
is extended over the whole region where $P_{\rm mf}(t) > P_{\rm mf}^{\rm appr}(t)$. For the constant background subtraction
$P_{\rm mf}^{\rm appr}(t) = P_{\rm mf}(0)$ this corresponds to a time window with $\tau \approx 0.8\,$s. The small increase
in $E_{\tau=0.01}$ in Table~\ref{t.recdata} with decreasing $\alpha$ might be interpreted as an indication of the emerging
Kelvin wave cascade, but since this increase is smaller than the uncertainty in the background subtraction, such a
conclusion on the basis of these calculations is unjustified.

\begin{figure}[!tb]
\centerline{
\includegraphics[width=0.98\linewidth]{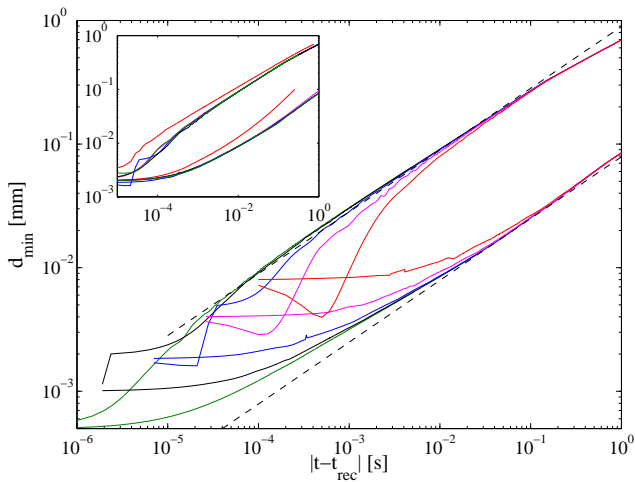}
}
\caption{(Color online)
Minimum distance $d_{min}$ between the reconnecting vortex segments. The lower group of curves plots $d_{min}$
before the reconnection, and the upper group of curves plots $d_{min}$ after the reconnection. \emph{Main panel:} $d_{min}$
calculated with five different resolutions when $\alpha$ = 0.01: $\Delta\xi_{\rm res}$ = 0.020\,mm (red), 0.010\,mm
(magenta), 0.005\,mm (blue), 0.0025\,mm (black), and 0.00125\,mm (green). The dashed lines are guides to the eye
such that for $t<t_{\rm rec}$ we have $d_{min}=\sqrt{\kappa(t_{\rm rec}-t)/16}$, while for $t>t_{\rm rec}$ we get
$d_{min}=\sqrt{8\kappa(t-t_{\rm rec})}$. In the \emph{inset} the results are shown for different $\alpha$ values
when the resolution $\Delta\xi_{\rm res}$ = 0.005\,mm: $\alpha = 1$ (red), 0.1 (magenta), 0.01 (blue), 0.001 (black),
and 0 (green).
}
\label{f.mindist}
\end{figure}

Our simulations also confirm the result from previous filament model calculations\cite{WaelePRL1994,AdachiJLTP2011,ZuccherFP2012,BoueLvov2013}
and from experimental observations\cite{BewleyPNAS2008reconnection,PaolettiPRL2008} that the minimum distance between
the reconnecting vortex segments behaves approximately as $d_{min} = C\sqrt{\kappa|t-t_{\rm rec}|}$, both before and after
the reconnection. This is illustrated in Fig.~\ref{f.mindist}. In our calculations the prefactor $C$ is about ten times
larger after the reconnection event than before, reflecting the larger curvatures appearing on the vortex. In classical
fluids and in Gross-Pitaevskii calculations the reconnection is found to be time asymmetric: $d_{min} = C|t-t_{\rm rec}|^\beta$,
where $\beta$ has different values before and after the reconnection.\cite{HussainPF2011,ZuccherFP2012}

\section{Discussion}

Our calculations have been performed on two reconnecting vortex rings with initial
radii of mainly  1\,mm. Since mutual friction affects every point on the vortex, on which the Kelvin waves
are rapidly distributed, the total energy must depend on the initial size of the rings. We have verified this
by repeating the calculations with rings of initial radii 0.1 and 10 mm (see upper inset of Fig.~\ref{f.Pmf0}).
In these cases the dissipated energy is about ten times smaller or larger, respectively. This is
consistent with the approximate scaling satisfied by Eq.~(\ref{e.vL}), as noted by Schwarz\cite{schwarz88}:
If lengths are scaled by a factor $\lambda$, times are scaled by a factor $\lambda^2$ and velocities are scaled by a factor $1/\lambda$, then the
motion is self-similar. Only the weakly changing logarithmic term (local term) breaks this scaling. This
means that the extra dissipated energy $E_{\rm rec}$, which we found to correspond to a few percent of the vortex length
with a ring radius of 1\,mm, can be considered to be independent of the initial radii of the reconnecting rings.

One may wonder whether the neglected length scales below our resolution limit might affect our conclusions.
This is unlikely. The resolution limit of the present calculations is two orders of magnitude below the
scale of the initial vortex configuration. The size of the vortex core, where the vortex filament model
has lost all its validity, is another two ($^3$He-B) to five ($^4$He) orders of magnitude below our
resolution limit. Finer resolutions were excluded owing to excessive computing times. In justifying our
results, we note that the calculations recover the expected large scale asymptote for $d_{min}$ in
Fig.~\ref{f.mindist}. Second, as seen in Table~\ref{t.recdata}, $E_{\rm rec}$ is largely independent of the
resolution limit $\Delta\xi_{\rm res}$. This statement can be made more quantitative by comparing the computed
results to the asymptotic dependence of the additional mutual friction power $P_{\rm mf}(t)-P_{\rm mf}(0)$ from
Fig.~\ref{f.Pmf0} integrated to include also the omitted small scale structures.
Such a comparison shows that the asymptotic value of $E_{\tau=0.01}$ is only a few percent larger than the
tabulated values.\cite{NoteAsymptotics} Based on these arguments about the large scale behavior, we argue
that the microscopic details of the reconnection and the exact small scale shape of the excitation kink do not alter the
qualitative features of our central conclusions. Moreover, since the decay time of Kelvin waves is proportional to $1/(\alpha k^2)$, the small but still finite mutual friction would quickly damp the waves with wave lengths smaller than our resolution.

The large enhancement of the mutual friction dissipation becomes understandable if one calculates the mutual
friction dissipation power $P_{\rm mf}$ using theoretically predicted Kelvin spectra. One may estimate that this
power diverges without a high-$k$ cut-off. \cite{Note1} 
At finite temperatures there must thus be a (temperature-dependent) mutual friction limited cut-off,
below which a steady-state Kelvin spectrum is realized.\cite{LvovEPL2012tempsupr} In our case the reconnection
kink excites Kelvin waves on all scales (limited by the resolution). Their spectrum can be argued to take a 
similar slope \cite{Note2} as the theoretically predicted steady-state spectrum\cite{KS2004,LN2010,SoninPRB2012}, 
and therefore dissipation is strongly increased. Recently a similar dissipation enhancement has been found in 
simulations considering a decaying vortex inside a cylinder.\cite{ZievePRB2010} These authors concluded that the 
larger dissipation at the lowest temperatures was due to a transfer of vortex length in consecutive reconnection 
events to surface-pinned remanent vortex loops trapped on the cylinder wall.

As seen in Fig.\,\ref{f.dissipEPA}, the reconnection causes a strong increase in energy dissipation,
whereas angular momentum dissipation is affected only weakly. Recent works on the propagating and precessing
vortex front in rotating superfluid $^3$He-B, both NMR measurements and vortex filament
calculations,\cite{HosioNC2013,HosioPRL2011} show that this feature might be valid more generally.
Thus, for isotropic vortex tangles, where the walls are far away and boundary effects can be neglected,
strong energy dissipation is possible and rapid decay in the total vortex length is observed. In contrast,
for polarized flow the slow removal of (angular) momentum limits the decay rate. In the vortex front measurements
the effects from pinning and surface friction were minimized by polishing the surfaces. In the zero-temperature
limit the axial front propagation velocity, which is related to energy dissipation, was observed to saturate at a 
value which is two orders of magnitude larger than the dissipation of angular momentum. In fact, it is possible
that the angular momentum dissipation might even vanish in these measurements in the limit of zero
temperature, if pinning or surface friction can be entirely excluded.

The weak dissipation of angular momentum might also explain why the spin down of a vortex array in a cylindrical
container of superfluid $^3$He-B is observed to be laminar.\cite{EltsovPRL2010}
Faster turbulent decay of energy requires rapid decay of angular momentum. If no additional torque
exists on vortices (which is the case in a cylindrically symmetric container with smooth walls),
then a faster decay of angular momentum is not possible. In contrast, in superfluid $^4$He pinning is
expected to be important, and this explains why the response is typically turbulent. Such a justification
is additionally supported by experiments and simulations on $^3$He-B, where an additional torque on
vortices was introduced by replacing part of the $^3$He-B with a layer of an $^3$He-A phase, in which
mutual friction is two orders of magnitude higher. In this case spin down was observed to become
turbulent.\cite{WalmsleyPRB2011}

Finally we note that we have not attempted to identify Kelvin modes quantitatively. Looking at
Fig.~\ref{f.confsT0}, one might argue that a reconnection event induces an inverse Kelvin wave cascade.
Kelvin waves immediately after the reconnection are at small scales, while somewhat later mainly waves
at long wavelengths have survived with large amplitude. This is consistent with the ideas presented by
Svistunov about the relaxation of the vortex angle.\cite{Svistunov1995}
Accurate identification of Kelvin waves on a nontrivial configuration of vortices is not a
simple task and can lead to unexpected errors.\cite{NiklasJLTP2013} This results from the fact that
generally Kelvin waves are not well defined, except in the case of a straight vortex or a simple planar
ring, and more sophisticated analysis is required.

\section{Conclusions}

Our simple case of two reconnecting vortex rings indicates that the Kelvin waves excited in a reconnection
event are efficient in dissipating energy, but are much less efficient in dissipating angular momentum.
The increased energy dissipation is not due to a Kelvin wave cascade, which transfers energy from large length
scales to small scales (owing to nonlinear coupling between Kelvin waves of different scales). Rather the enhanced
dissipation is caused by the direct transfer of energy from the excited Kelvin waves to the normal fluid by mutual
friction damping.

Our numerical estimates for the extra mutual friction dissipation indicate that a single reconnection event can
dissipate energy to an amount which in our example case corresponds to a few percent of the vortex length, an
unexpectedly large value which additionally is almost temperature independent.
In turbulent vortex tangles dissipation per reconnection event is expected to be smaller,
because it is likely limited by the time between reconnections, the inter-vortex distance, vortex polarization,
etc. At still lower temperatures than considered here, where $\alpha \ll 10^{-3}$, for mutual friction
to dissipate all the reconnection energy, the required time diverges. In this limit the Kelvin wave cascade
might become a more dominant dissipation mechanism, as believed today. \\

\begin{acknowledgments}
This work is supported by the Academy of Finland (Grant No. 218211)
and in part by the European Union 7th Framework Programme (FP7/2007-2013, Grant No. 228464 Microkelvin)
and by the Academy of Finland through its LTQ CoE grant (Project No. 250280).
I thank V.B. Eltsov, M. Krusius, N. Hietala, and G.E. Volovik, for their comments and suggestions,
and the CSC - IT Center for Science, Ltd., for the allocation of computational resources.
\end{acknowledgments}



\begin{thebibliography}{99}


\bibitem{Svistunov1995}
B.~V. Svistunov, Phys. Rev. B {\bf 52}, 3647-3653 (1995).

\bibitem{VinenJLTP2002}
W.~F. Vinen and J.~J. Niemela, J. Low Temp. Phys. {\bf 128}, 167-231 (2002).

\bibitem{BradleyPRL2006decay}
D.~I. Bradley, D.~O. Clubb, S.~N. Fisher, A.~M. Gu\'enault, R.~P. Haley, C.~J. Matthews, G.~R. Pickett, V. Tsepelin, and K. Zaki,
Phys. Rev. Lett. {\bf 96}, 035301 (2006); D.~I. Bradley, S.~N. Fisher, A.~M. Gu\'enault, R.~P. Haley, G.~R. Pickett, D. Potts, and V. Tsepelin, Nature Phys. {\bf 7}, 473 (2011).

\bibitem{FrontPRL2007} V.~B. Eltsov, A.~I. Golov, R. de~Graaf, R. H\"anninen, M. Krusius, V.~S. L'vov, and R.~E. Solntsev, Phys. Rev. Lett. \textbf{99}, 265301 (2007).

\bibitem{WalmsleyPRL2007}
P.~M. Walmsley, A.~I. Golov, H.~E. Hall, A.~A. Levchenko, and W.~F. Vinen,
Phys. Rev. Lett. {\bf 99}, 265302 (2007).

\bibitem{NemirovskiiPRB2010} S.~K Nemirovskii, Phys. Rev. B {\bf 81}, 064512 (2010).

\bibitem{KS2004}
E. Kozik and B. Svistunov, Phys. Rev. Lett. {\bf 92}, 035301 (2004).

\bibitem{LN2010}
V.~S. L'vov and S. Nazarenko, JETP Lett. {\bf 91}, 428-434 (2010).

\bibitem{SoninPRB2012}
E.~B. Sonin, Phys. Rev. B {\bf 85}, 104516 (2012).

\bibitem{KursaPRB2011}
M. Kursa, K. Bajer, and T. Lipniacki, Phys. Rev. B {\bf 83}, 014515 (2011).

\bibitem{KerrPRL2011} R.~M. Kerr, Phys. Rev. Lett. {\bf 106}, 224501 (2011).

\bibitem{BarenghiPRL2002} C.~F. Barenghi, and D.~C. Samuels, Phys. Rev. Lett. {\bf 89}, 155302 (2002).

\bibitem{NemirovskiiPRB2012} L. Kondaurova and S.K. Nemirovskii, Phys. Rev. B {\bf 86}, 134506, (2012).

\bibitem{KivotidesEPL2001} D. Kivotides, C.~F. Barenghi, and D.~C. Samuels, Europhys. Lett. {\bf 54} 774-778 (2001).

\bibitem{HosioNC2013}
J.~J. Hosio, V.~B. Eltsov, P.~J. Heikkinen, R. H\"anninen, M. Krusius, and V.~S. L'vov,
Nat. Commun. {\bf 4}, 1614 (2013). 

\bibitem{HosioPRL2011}
J.~J. Hosio, V.~B. Eltsov, R. de~Graaf, P.~J. Heikkinen, R. H\"anninen, M. Krusius, V.~S. L'vov, and G.~E. Volovik,
Phys. Rev. Lett. {\bf 107}, 135302 (2011). 

\bibitem{schwarz85}
K.~W. Schwarz, Phys. Rev. B {\bf 31}, 5782-5804 (1985).

\bibitem{SaffmanBook}
P.~G. Saffman, \emph{Vortex dynamics}, (Cambridge University Press, Cambridge, England, 1992).

\bibitem{Alekseenko2007}
For a more detailed definition for $P$ and $A$ see e.g. S.~V. Alekseenko, P.~A. Kuibin, and V.~L. Okulov,
\emph{Theory of Concentrated Vortices}, (Springer-Verlag, Berlin, 2007).

\bibitem{KoplikLevinePRL1993} J. Koplik, and H. Levine, Phys. Rev. Lett. {\bf 71} 1375 (1993).

\bibitem{BewleyPNAS2008reconnection}
G.~P. Bewley, M.~S. Paoletti, K.~R. Sreenivasan, and D.~P. Lathrop, Proc. Natl. Acad. Sci. U.S.A. {\bf 105}, 13707-13710 (2008).

\bibitem{TsubotaPRB2000}
M. Tsubota, T. Araki, S.~K. Nemirovskii, Phys. Rev. B {\bf 62}, 11751-11762 (2000).

\bibitem{KondaurovaJLTP2008}
L.~P. Kondaurova, V.~A. Andryuschenko, and S.~K. Nemirovskii, J. Low Temp. Phys. {\bf 150}, 415-419 (2008).

\bibitem{BaggaleyJLTP2012rec}
A.~W. Baggaley, J. Low Temp. Phys. {\bf 168}, 18-30 (2012).

\bibitem{KondaurovaLvov2013}
L. Kondaurova, V. L'vov, A. Pomyalov, and I. Procaccia, arXiv:1306.6167 (2013).

\bibitem{BaggaleyPRB2011spectrum}
A.~W. Baggaley, and C.~F. Barenghi, Phys. Rev. B {\bf 83}, 134509 (2011).

\bibitem{KivotidesPRL2001} D. Kivotides, J.~C. Vassilicos, D.~C. Samuels, and C.~F. Barenghi,
Phys. Rev. Lett. {\bf 86}, 3080-3083 (2001).

\bibitem{LeadbeaterPRL2001} M. Leadbeater, T. Winiecki, D.~C. Samuels, C.~F. Barenghi, and C.~S. Adams,
Phys. Rev. Lett. {\bf 86}, 1410-1413 (2001).

\bibitem{WaelePRL1994}
A.~T.~A.~M. de~Waele, and R.~G.~K.~M. Aarts, Phys. Rev. Lett. {\bf 72}, 482-485 (1994).

\bibitem{AdachiJLTP2011}
M. Tsubota, and H. Adachi, J. Low Temp. Phys. {\bf 162}, 367-374 (2011).

\bibitem{ZuccherFP2012}
S. Zuccher, M. Caliari, A.~W. Baggaley, and C.~F. Barenghi, Phys. Fluids {\bf 24}, 125108 (2012).

\bibitem{BoueLvov2013}
L. Bou\'e, D. Khomenko, V.S. L'vov, and I. Procaccia, arXiv:1307.5282 (2013).

\bibitem{PaolettiPRL2008}
M.~S. Paoletti, M.~E. Fisher, K.~R. Sreenivasan, and D.~P. Lathrop, Phys. Rev. Lett. {\bf 101}, 154501 (2008).

\bibitem{HussainPF2011}
F. Hussain and K. Duraisamy, Phys. Fluids {\bf 23}, 021701, (2011).

\bibitem{schwarz88}
K.~W. Schwarz, Phys. Rev. B {\bf 38}, 2398-2417 (1988).

\bibitem{NoteAsymptotics}
The asymptotic form for $E_\tau/\rho_s\kappa^2 = \alpha A \int_{t=0}^{\tau/\alpha} (\alpha t)^{-\gamma}dt$,
where the fitted values $A$ $\approx$ 0.491 and $\gamma$ $\approx$ 0.581 are obtained when $\alpha$ = 0.01.
This convergent integral gives the right-hand side in millimeters when the times are measured in seconds, and is
insensitive to the actual value of the core size, $a_0$, even if a cutoff is introduced.

\bibitem{Note1}
For a straight vortex along the $z$-axis with Kelvin waves, whose configuration is given by $w(z)=x(z)+iy(z)$,
the Kelvin spectrum is given by $n_k = |w_k|^2$, where $w_k$ is the Fourier transformation of $w(z)$. For a Kelvin
spectrum $n_k = k^{-\eta}$ the localized induction approximation states that in the limit $k_{\rm max} \sim 1/a_0 \rightarrow \infty$,
the mutual friction dissipation power $P_{\rm mf}$ diverges if $\eta \leq 5$.

\bibitem{LvovEPL2012tempsupr} L. Bou\'e, V. L'vov, and I. Procaccia, Europhys. Lett. {\bf 99}, 46003 (2012).

\bibitem{Note2}
The exact spectrum depends on the form of the reconnection cusp. Assuming a discontinuous derivate, several kinds of kinks produce a spectrum $n_k = |w_k|^2 \propto k^{-\eta}$ with $\eta \leq 4$.
Our preliminary simulations, which follow the pre-reconnection dynamics of two
(initially straight and perpendicular) line vortices, produce a spectrum with $\eta \approx 4$.

\bibitem{ZievePRB2010} I.~H. Neumann, and R.~J. Zieve, Phys. Rev. B {\bf 81}, 174515 (2010).

\bibitem{EltsovPRL2010}
V.~B. Eltsov, R. de~Graaf, P.~J. Heikkinen, J.~J. Hosio, R. H\"anninen, M. Krusius, V.~S. L'vov,
Phys. Rev. Lett. {\bf 105}, 125301 (2010). 

\bibitem{WalmsleyPRB2011}
P.~M. Walmsley, V.~B. Eltsov, P.~J. Heikkinen, J.~J. Hosio, R. H\"anninen, and M. Krusius,
Phys. Rev. B {\bf 84}, 184532 (2011). 

\bibitem{NiklasJLTP2013}
R. H\"anninen, and N. Hietala, J. Low Temp. Phys. {\bf 171}, 485-496 (2013).


\end{thebibliography}
\end{document}